\documentclass[10pt]{dis03}
\usepackage{epsf,amsmath}

\usepackage{graphics}
\usepackage{epsfig}
\usepackage{my}
\usepackage[latin1]{inputenc}

\newcommand{\qb}{\ensuremath{\bar{q}} }

\def\prp{\perp}

\def\kt{\ensuremath{k_\prp}}

\def\qt{\ensuremath{q_\prp}}

\def\sub#1{\ensuremath{_{\mbox{\scriptsize #1}}}}
\def\alb{\ensuremath{\bar{\alpha}\sub{S}}}

\newcommand{\alphasb}{\alb}

\newcommand{\asb}{\alb}

\def\CASCADE{{\sc Cascade}}

\newcommand{\Pmax}{\bar{q}}

\begin{document}

\title{Status of CCFM - un-integrated gluon  densities}

\author{M. Hansson and H. Jung \\
Department of Physics, Lund University,\\
Lund,  Sweden \\
E-mail:hansson@mail.desy.de and jung@mail.desy.de}

\maketitle

\begin{abstract}
\noindent New fits of the unintegrated gluon density 
obtained from CCFM evolution to HERA $F_2(x,Q^2)$ data are presented. Also
predictions of the unintegrated gluon density of the real photon are presented.
\end{abstract}

\section{The CCFM splitting function} 
A general review on small $x$ physics and CCFM evolution 
can be found in~\cite{smallx_2001}. 
The original CCFM splitting function is given by :
\begin{equation}
P_{gg}(z,\qb,\kt )= \frac{\alphasb(\qb^2)}{1-z} + 
\frac{\alphasb(\kt^2)}{z} \Delta_{ns}(z,\qb^2,\kt)
\label{Pgg}
\end{equation}
with $\bar{q} = q(1-z)$ and 
with the non-Sudakov form factor $\Delta_{ns}$ defined as:
\begin{eqnarray}
\log\Delta_{ns}(z_i,\qb^2,\kt)& = & -\alphasb
                  \int_{z}^1 \frac{dz'}{z'} 
                        \int \frac{d q^2}{q^2}  \cdot 
                  \Theta(\kt-q)\Theta(q-z'\qb)
                      \nonumber   \\
	    & = &  -\alphasb
                  \int_{z}^1 \frac{dz'}{z'} 
                        \int^{\kt^2}_{(z'\qb)^2} \frac{d q^2}{q^2} 
               \label{non_sudakov_int}       
\end{eqnarray}
Here only the singular terms $1/z$ and $1/(1-z)$ were included and for
simplicity the scale in the running $\alpha_s$ was not treated in the same way
for the small and large $z$ part.
\par
Due to the angular ordering 
a kind of random walk in the propagator gluon 
\kt can be performed.
For values of $\kt < \kt^{cut}$ the non-perturbative region is entered, which is
avoided in a strictly \qt-ordered evolution (DGLAP). 
The region of small \kt\ 
is characterized by $\alpha_s$ and the parton density  being large, 
and collective phenomena, like gluon recombination or saturation might play a
role. 
At such small \kt\,  the total cross section is expected to rise only weakly with
energy, equivalently to a constant $xG(x,Q)$ for small $x$ and $Q$.    
A practical treatment is
therefore to fix $\alpha_s(\mu)$ to 
$\alpha_s(Q_0) \sim 0.6$ for $\mu < Q_0 $. Until $\kt > \kt^{cut}$ is reached, 
no gluon emissions are allowed, but energy-momentum 
conservation is properly treated.
In the {\bf JS} unintegrated gluon density \cite{jung_salam_2000}, the soft
region was defined by $\kt^{cut} = 0.25$~GeV.
In the new sets presented here, $\kt^{cut} = Q_0$ was
chosen, with $Q_0$ being the collinear cut for the real emissions (the scale for
resolvable branchings). 

\par
Following the arguments in \cite{smallx_2001}, it was investigated in
\cite{jung-dis02} to change the scale in $\alpha_s$  to $q^2(1-z)^2$ everywhere, 
in $1/z$ part of the splitting function as well as in the non-Sudakov 
form factor. 
From eq.(\ref{non_sudakov_int}) it is obvious, 
that a special treatment of the soft region (and the lower integration limit) 
is needed, because $q'$ can become very small and even 
$q' < \Lambda_{qcd}$  at small values of $z'$. 
The problematic region in the non-Sudakov form factor in 
eq.(\ref{non_sudakov_int}) is avoided by fixing $\alpha_s(\mu)$ for 
$\mu<0.9$ GeV~\cite{smallx_2001,jung-dis02}:
\begin{eqnarray}
\log\Delta_{ns}  &              = &-C_a \cdot \int_{z_1}^{z_0} \frac{dz'}{z'}
                  \int^{k_{t}^2}_{(z'\qb)^2} 
			\frac{d q^2}{q^2}
                  \frac{1}{\log(q/\Lambda_{qcd})} \nonumber \\
		& & -\asb(q_{cut}) \int_{z}^{z_c} \frac{dz'}{z'}
                 \int^{k_{t}^2}_{(z_c\qb)^2} 
			\frac{d q^2}{q^2} \Theta(z_c - z)
		\label{non_sudakov_asq} 	
\end{eqnarray}
with $ C_a = 36/(33-2n_f)$ and $n_f$ being the number of active flavours. The
integration limits are defined by
$z_c=q_{cut}/\qb$, $q_{cut}=0.9$ GeV and $z_1 =\max(z,z_c)$.  
However in practical application we observe only a
small effect from changing the scale of the small $z$ part from \kt to \qt.
\par
At very high energies, the $1/z$ term in $P_{gg}$
will certainly be dominant. However, at present colliders
the non-singular terms, as suggested in  
\cite{smallx_2001} should also be included:
\begin{equation}
P_{gg}(z,\qb,\kt ) = \asb \left(\kt^2 \right) 
\left( \frac{(1-z)}{z} + \frac{z(1-z)}{2}\right) \Delta_{ns} 
  + \asb (\qb^2) \left(\frac{z}{1-z} + 
\frac{z(1-z)}{2}\right) .
\label{fullsplitt}
\end{equation}
\par
As already mentioned before, the change of the scale in $\alpha_s$ from \kt to
\qt does not produce significant differences. For simplicity,   
\kt is still taken as the scale for the small $z$ part
in the non-Sudakov form factor.
\section{Unintegrated gluon density of the Proton} 
\begin{figure}[htb]
\centerline{\rotatebox{0.}{\scalebox{0.32}{\includegraphics{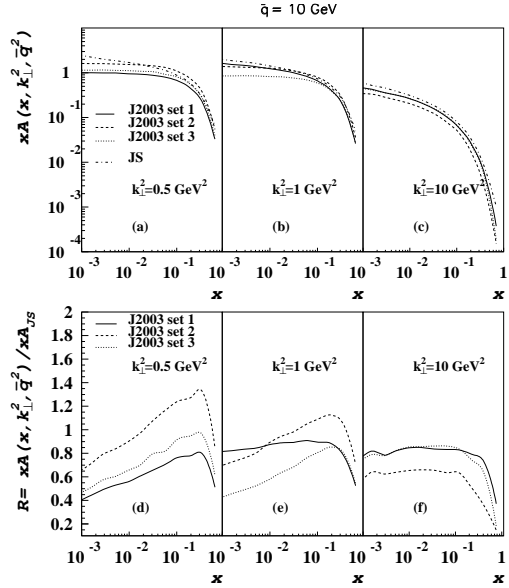}}}}
 \caption[*]{
 {\it Comparison of the different sets of unintegrated gluon densities obtained
 from the CCFM evolution as described in the text. In $(a-c)$ the unintegrated
 gluon density is shown as a function of $x$ for different values of \kt\ at a
 scale of $\bar{q}=10$ GeV. In $(d-f)$ the ratio 
 $R=\frac{x{\cal A}(x,\kt^2,\Pmax^2)}
 {x{\cal A}(x,\kt^2,\Pmax^2)_{\bf JS}}$ as a function of $x$ 
 for different values of \kt\  is shown.
 \label{ccfm-new}}}
\end{figure}
The CCFM evolution equations have been solved 
numerically~\cite{jung_salam_2000} using a Monte Carlo method. 
Three new sets ({\bf J2003 set 1 - 3}, details are given in Tab.~\ref{pdfsets}) of
unintegrated gluon densities were determined and compared to the
previous one {\bf JS} ~\cite{jung_salam_2000}. The 
input parameters were fitted to
describe the structure function $F_2(x,Q^2)$ 
in the range $x < 5 \cdot 10^{-3}$ and  $Q^2 > 4.5$~GeV$^2$ 
as measured at H1~\cite{H1_F2_1996,H1_F2_2001} and  
ZEUS~\cite{ZEUS_F2_1996,ZEUS_F2_2001}. 
Set {\bf JS}~\cite{jung_salam_2000} was fitted only to 
$F_2(x,Q^2)$ of Ref.~\cite{H1_F2_1996}.
\begin{table}[htb]
\begin{center}
\begin{tabular}{| c | c | c | c | c | }
\hline
\hline
set & $P_{gg}$ & $Q_0$   & $\kt^{cut}$ (GeV) & $\chi^2/ndf$ \\
\hline
{\bf JS}~\protect\cite{jung_salam_2000}          
                  &  eq.(\ref{Pgg},\ref{non_sudakov_int})  
			& 1.40  GeV &  0.25  & 1197/248 = 4.8\\
{\bf J2003 set 1} &  eq.(\ref{Pgg},\ref{non_sudakov_int})  
                  & 1.33 GeV & 1.33 & 321/248 = 1.29\\
{\bf J2003 set 2} &  eq.(\ref{fullsplitt})  
                  & 1.18 GeV &  1.18  & 293/248 = 1.18\\
{\bf J2003 set 3} &  eq.(\ref{non_sudakov_asq})  
                  & 1.35 GeV & 1.35  & 455/248 = 1.83 \\
\hline
\hline
\end{tabular}
\caption{The different settings of the CCFM unintegrated gluon densities. In 
{\bf J2003 set 2} and {\bf J2003 set 3} the lower integration limit  
in the non-Sudakov form
factor is changed from eq.(\ref{non_sudakov_int}) to
$((z_c q)^2 )$.
In the last column, the $\chi^2/ndf$ to HERA 
$F_2$ data~\protect\cite{H1_F2_1996,H1_F2_2001,ZEUS_F2_1996,ZEUS_F2_2001}
 is given (for $x< 5\cdot 10^{-3}$ and $Q^2 > 4.5 $ GeV$^2$). }
\label{pdfsets}
\end{center}
\end{table}
A comparison of the different sets of CCFM unintegrated gluon densities is shown
in Fig.~\ref{ccfm-new}. It is clearly seen, that the treatment of the soft
region, defined by $\kt < \kt^{cut}$ influences the behavior at small $x$ and
small \kt. It is interesting to note, that the change of the scale form \kt\ to
\qt\  in {\bf J2003 set 2} is  visible only in the small \kt\ region, whereas
at larger \kt\ {\bf J2003 set 1} and {\bf J2003 set 3} agree reasonably well. 
In Tab.~\ref{pdfsets} the parameters of the CCFM unintegrated gluon densities
are summarized and also the 
$\chi^2$ of the fits are shown.

\section{Unintegrated gluon density of the Photon} 
\begin{figure}[htb]
\centerline{\rotatebox{0.}{\scalebox{0.32}{\includegraphics{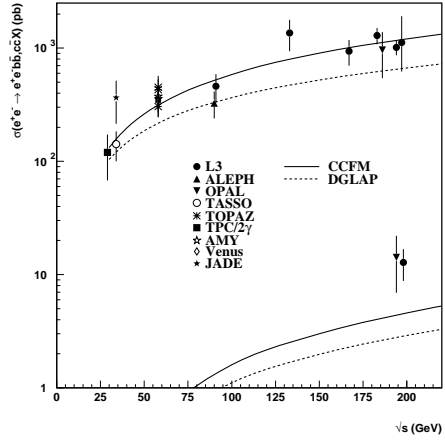}}}}
 \caption[*]{{\it The cross section for heavy quark production obtained with \CASCADE\  using
 the CCFM unintegrated gluon density of the photon compared to measurements in $\gamma
 \gamma$ collisions.
 \label{bbar_lep}}}
\end{figure}

We use the same parameter settings as given in Tab.~\ref{pdfsets}, to calculate the CCFM
unintegrated gluon density of the real photon. 
For the input gluon density at the scale $Q_0$
we  chose GRV set~\cite{GRVgam},
 which also determined the normalization. The set corresponding to 
 {\bf J2003 set 2} is  used in \CASCADE ~1.2~\cite{CASCADEMC} 
to calculate heavy quark production in $\gamma \gamma$ reactions  
and are compared with the measurements at LEP
(Fig.~\ref{bbar_lep}). Charm production is reasonably well described, 
whereas bottom production falls below the measurement. 
The prediction obtained in 
\kt-factorization is
only slightly larger than that obtained in the collinear approach at NLO. 

\section*{Acknowledgments} 
We are grateful to the organizers of the DIS2003 workshop for the 
stimulating atmosphere
and the good organization.

\end{document}